\documentclass[11pt]{iopart}
\usepackage{graphicx}
\usepackage{gensymb}
\usepackage[style=ieee,citestyle=numeric-comp,sorting=none]{biblatex}
\usepackage{xcolor}
\usepackage[export]{adjustbox}
\usepackage{subcaption}
\begin{document}
\title{Bayesian parameter estimation for characterising mobile ion vacancies in perovskite solar cells}
\author{Samuel McCallum$^1$, Oliver Nicholls$^1$, Kjeld Jensen$^2$, Mathew V. Cowley$^3$, James Lerpini\`{e}re$^1$, Alison B Walker$^1$}
\address{$^1$Department of Physics, University of Bath, Bath, UK}
\address{$^2$Applied Research, British Telecommunications, UK}
\address{$^3$Institute for Sustainability, University of Bath, Bath, UK}
\ead{pysabw@bath.ac.uk}
\date{}
\begin{abstract}
To overcome the challenges associated with poor temporal stability of perovskite solar cells, methods are required that allow for fast iteration of fabrication and characterisation, such that optimal device performance and stability may be actively pursued. Currently, establishing the causes of underperformance is both complex and time-consuming, and optimisation of device fabrication thus inherently slow. Here, we present a means of computational device characterisation of mobile halide ion parameters from room temperature current-voltage (J-V) measurements \emph{only}, requiring $\sim 2$ hours of computation on basic computing resources. With our approach, the physical parameters of the device may be reverse modelled from experimental J-V measurements. In a drift-diffusion model, the set of coupled drift-diffusion partial differential equations cannot be inverted explicitly, so a method for inverting the drift-diffusion simulation is required. We show how Bayesian Parameter Estimation (BPE) coupled with a drift-diffusion perovskite solar cell model can determine the extent to which device parameters affect performance measured by J-V characteristics. Our method is demonstrated by investigating the extent to which device performance is influenced by mobile halide ions for a specific fabricated device. The ion vacancy density $N_0$ and diffusion coefficient $D_I$ were found to be precisely characterised for both simulated and fabricated devices. This result opens up the possibility of pinpointing origins of degradation by finding which parameters most influence device J-V curves as the cell degrades.
\end{abstract}
\maketitle
\section{Introduction}
Lead-halide perovskites (LHPs) have attracted intense research activity as light harvesting layers in solar cell devices due to power conversion efficiency (PCE) increases of 9-25.8$\%$ since 2009 \cite{Sharif23}, rivalling the record efficiency reported for silicon-based solar cells of 26.7$\%$ \cite{Green23}. Additionally, perovskite solar cells (PSCs) may be fabricated using low-energy, low-cost manufacturing  techniques such as solution processing \cite{Razza16}. Despite the progress to large power conversion efficiencies, problems with the reproducibility and stability of PSCs remains \cite{Goetz22,Boyd19}. The decrease of PCE over time originates in extrinsic factors such as exposure to heat \cite{Kim17}, light \cite{Wang19}, humidity \cite{Leguy15} and oxygen \cite{Fang16} and intrinsic factors such as the structural stability of the perovskite \cite{Han18} and migration of mobile ion vacancies across the perovskite layer \cite{Eames15}. 

To find strategies and fabrication techniques in addressing these problems, a large number of parameters is required to  specify layer widths and materials parameters for the perovskite layer and the electron and hole transport layers (ETL, HTL respectively). These parameters are often not accurately known and may vary from device to device. Identifying which of these parameters is sensitive to the fabrication method and extent of degradation of a \emph{specific} fabricated device gives important insights on the origins of subpar device performance. The PSCs can be characterised by experimental measurements of, for example, current-voltage (J-V) characteristics at varying voltage scan rates. These characteristics do not however give directly the device and materials parameter values. Here we present a method for reverse modelling, namely mapping from J-V device measurements to device and materials parameters.  We do so by solving the inverse parameter problem where a drift-diffusion (DD) simulation is combined with Bayesian parameter estimation, BPE. It is not possible to invert explicitly the coupled partial differential equations in the DD model.

\begin{figure}
\centering
\begin{subfigure}{.5\textwidth}
  \centering
  \includegraphics[width=0.8\linewidth]{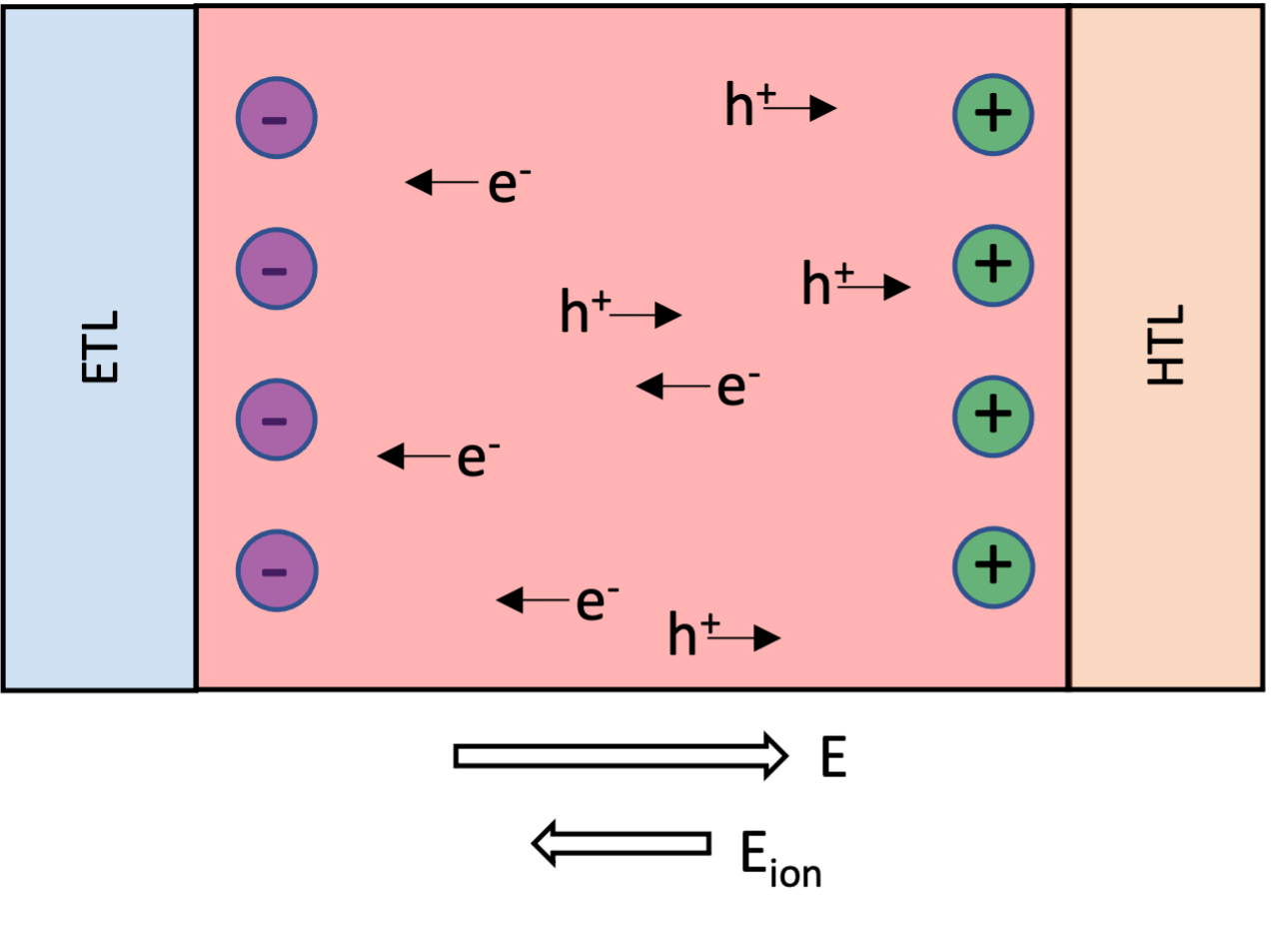}
  \caption{}
\end{subfigure}%
\begin{subfigure}{.5\textwidth}
  \centering
  \includegraphics[width=\linewidth]{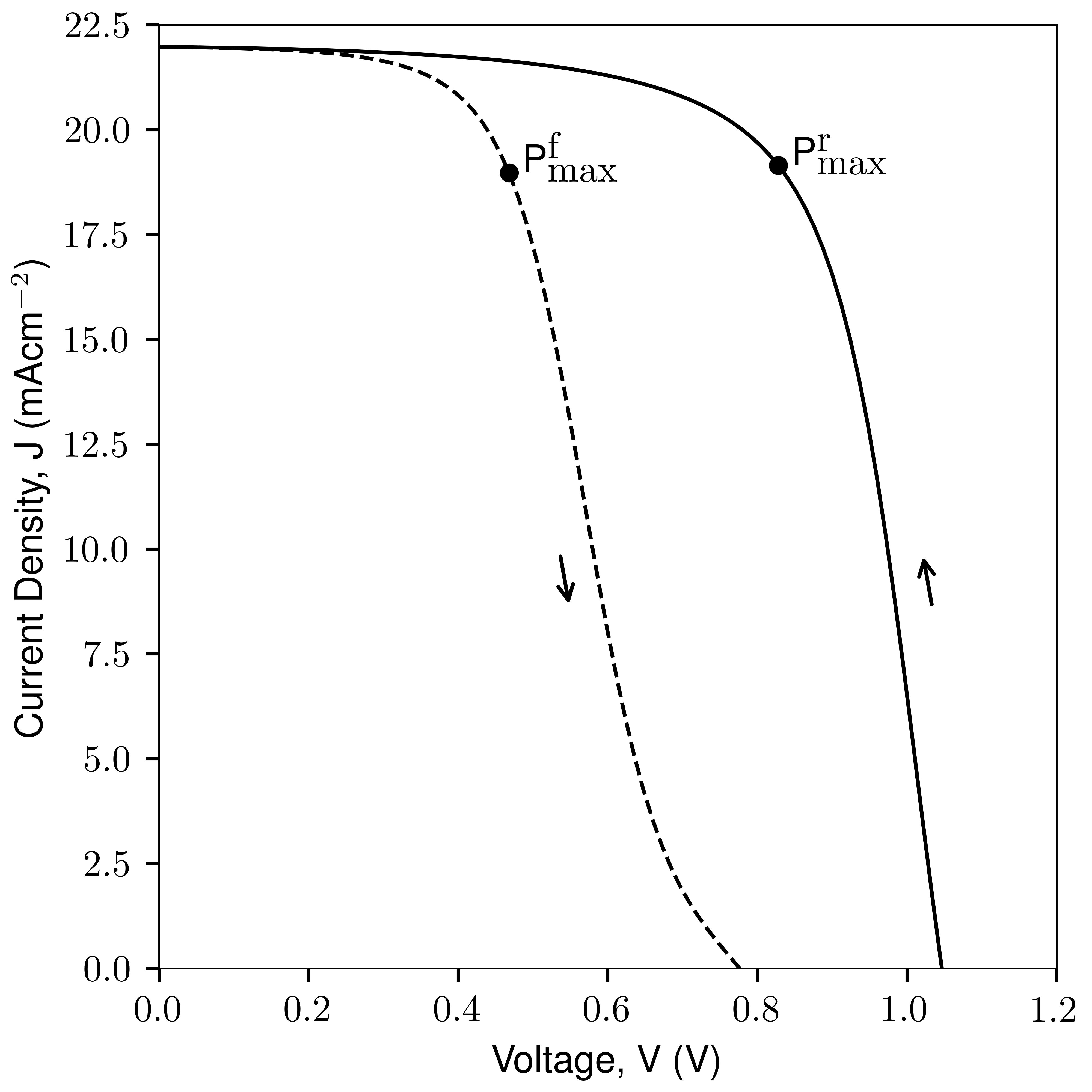}
  \caption{}
\end{subfigure}
\caption{(a) Illustration of halide ion vacancy migration in perovskite solar cells. Mobile halide ions (purple) migrate through ion vacancies (green) to the electron transport layer (ETL) boundary, producing an electric field E$_{\mathrm{ion}}$. This field partially screens the electric field due to V$_\mathrm{bi}$ and V$_\mathrm{app}$. Electrons and holes are indicated by e$^-$ and h$^+$, respectively. (b) Simulated example of J-V hysteresis where the current density on the forward voltage-scan (dashed line), differs from the reverse voltage-scan (solid line); arrows indicate voltage-scan direction. The maximum power point (P$_\mathrm{max}$) for both the reverse (superscript $\mathrm{r}$) and forward (superscript $\mathrm{f}$) scan are shown, indicating the operating power losses as a result of hysteresis.}
\label{fig:ion_migration}
\end{figure}

Despite strong evidence for the migration of halide ions, and an understanding of its effect on device performance, it is not clear how to determine whether performance losses in a specific device may be attributed to ion migration. For example, a recent study has been made of degradation pathways linked to iodide diffusion and iodide reactions. \cite{Bitton23} The electric field in the perovskite layer results from the built in voltage of the device V$_\mathrm{bi}$ and the externally applied voltage V$_\mathrm{app}$, such that the electric field across the perovskite $E=(\mathrm{V_{bi}}-\mathrm{V_{app}})/b_p$ where $b_p$ is the width of the perovskite layer, once potential drops across the charge transport layers are allowed for. The electric field causes the mobile ion vacancies to migrate across the perovskite layer, resulting in a build up of negatively charged ions at the boundary between the perovskite and ETL and positively charged ion vacancies at the HTL (see Figure \ref{fig:ion_migration}). In turn, the resulting non-uniform charge distribution partially screens the total electric field in the perovskite layer - severely affecting charge transport, reducing the power output of the device and causing current-voltage hysteresis \cite{GRodriguez22, LeCorre22, Thiesbrummel21}. The effect of halide ion migration can be detrimental to cell performance, causing significant reduction in PCE over time on the order of seconds to days \cite{Domanski18}. 

As shown by density functional theory (DFT) calculations \cite{Eames15, Walsh15}, halide ions (e.g. iodide I$^-$) are the most likely candidate to undergo migration through vacancy pathways due to the ions' large diffusion coefficients in comparison to other ionic species. Migration of halide ions within the perovskite layer has also been confirmed experimentally \cite{deQuilettes16, Minns17}. J-V characteristics such as hysteresis are strongly dependent on mobile ion vacancy parameters, the mobile ion vacancy density $N_0$ and mobile ion diffusion coefficient $D_I$ \cite{Courtier19EES, Cave20}. Here, we demonstrate our reverse modelling approach by mapping from J-V device measurement to ion vacancy parameters. Our method characterises mobile ion vacancy parameters ($N_0$ and $D_I$) from room temperature J-V device measurements \emph{only}, without further experiment. Knowledge of  $N_0$ and $D_I$ for a specific device can then be used to minimise the effect of ion migration on the performance and stability of PSCs, such as optimising experimental processing \cite{Burschka13, Singh18}, tailoring perovskite composition \cite{Pering22, Kamat21}, and tuning transport layer properties \cite{Wojciechowski15, Courtier19EES}, can be tested. Without appropriate characterisation of mobile ion vacancy parameters for a specific device, it is difficult to analyse how the reduction in performance, potentially caused by mobile ion vacancies, depends on processing and fabrication methods. This limits the rate at which experimental optimisation can be achieved and restricts progress towards improving the temporal stability of PSCs.

In our approach, BPE \cite{Bretthorst88, Christensen01, Thiemann01} was used to iteratively query a drift-diffusion PSC model to obtain a posterior distribution over the ion vacancy parameters most likely to reproduce experimental J-V curves. BPE has previously been used in solar cell applications to characterise the charge carrier mobility, minority carrier lifetime and conduction and valence band offsets between the light-harvesting and transport layers \cite{Brandt17}. Kirchartz et al. demonstrated that BPE may be applied to PSCs and also utilised a neural network model to approximate the drift-diffusion simulations \cite{Kirchartz23}. However, the effect of ion migration present in PSCs is not included in reference \cite{Kirchartz23}. Our choice of BPE method requires only single temperature J-V measurements at two voltage scan-rates, in comparison to the multiple temperature \cite{Brandt17} and illumination \cite{Kirchartz23} measurements used in previous studies. We model both charge carriers and mobile ion vacancies, and allow all parameters associated with the device (perovskite and transport layer parameters, see Table 1) to influence performance. Additionally, we determine posterior distributions over all fundamental device parameters, allowing for a greater understanding of the underlying physical processes affecting performance. 

\section{Methods}
Solving the inverse problem of mapping from J-V device measurement to device parameters requires a model of the forward relationship (device parameters to J-V measurement) which may then be inverted. Here, this forward relationship has been approximated by utilising a coupled electron-hole-ion drift diffusion device model, namely IonMonger \cite{Courtier18, Courtier19JCE}. IonMonger is capable of simulating the current across the device as a function of the experimental conditions (illumination, applied bias, voltage scan-rate) and fundamental physical parameters (see Table 1).  We propose a Bayesian Parameter Estimation (BPE) approach to solve the inverse problem; the advantages and disadvantages of this method are discussed in section 3.

BPE applies Bayes' Rule to inversely derive a posterior distribution $p(\theta | \mathbf{y})$ over a set of parameters $\theta$ associated with a measured outcome $\mathbf{y}$,
\begin{equation}
    p(\theta | \mathbf{y}) = \frac{p(\mathbf{y} | \theta) p(\theta)}{\int_{\theta} p(\mathbf{y} | \theta) p(\theta)}.
\end{equation}
$p(\mathbf{y} | \theta)$ is the likelihood of obtaining the measured outcome $\mathbf{y}$ from $\theta$, $p(\theta)$ is the prior distribution over $\theta$, and $\int_{\theta} p(\mathbf{y} | \theta) p(\theta)$ is a normalising factor. In this application, $\theta$ is the set of fundamental physical device parameters associated with a PSC device (see Table 1) and $\mathbf{y}$ is the set of characteristics obtained from a measured J-V curve; specifically $\mathbf{y} = [J_\mathrm{sc}, V_\mathrm{oc}^\mathrm{r}, V_\mathrm{oc}^\mathrm{f}, \mathrm{P}_\mathrm{max}^\mathrm{r}, \mathrm{P}_\mathrm{max}^\mathrm{f}]$, where $J_\mathrm{sc}$ is the short-circuit current density, $V_\mathrm{oc}$ is the open-circuit voltage, and the superscripts $\mathrm{r}$ and $\mathrm{f}$ indicate quantities measured on the reverse and forward scans, respectively.

The prior distribution over perovskite parameters $p(\theta)$ was assumed to be a uniform distribution over each parameter, representing the inherent epistemic uncertainty prior to any measurement of the device. Table 1 lists an estimated parameter range within each prior distribution for the TiO$_2$-MAPbI$_3$-Spiro PSC architecture.

The likelihood $p(\mathbf{y} | \theta)$ of obtaining a measured set of J-V characteristics $\mathbf{y}$ from a set of parameters $\theta$ was determined by querying IonMonger. Let the prediction of $\mathbf{y}$ by IonMonger be $\mathbf{\tilde{y}}$ for given $\theta$. The likelihood that $\theta$ produces the observed characteristics $\mathbf{y}$ can therefore be modelled as a multivariate Gaussian distribution with mean $\mathbf{y}$,
\begin{equation}
    p(\mathbf{y} | \theta) = \mathcal{N}(\mathbf{\tilde{y}}; \mu=\mathbf{y}, \Sigma=\mathbf{\sigma}^2I),
\end{equation}
where the covariance is determined by the set of experimental uncertainties in each measured J-V characteristic $\mathbf{\sigma}$, assumed to be $\pm 0.1\%$. The dominating uncertainty originates from measurement of the current density, as shown by Brandt et al. \cite{Brandt17} to be of the order $\pm 0.1\%$, and $\sigma$ was therefore set to this value. Note that the covariance matrix is diagonal, quantifying an assumption that the measured J-V characteristics for a specific device are independently distributed.

The posterior distribution $p(\theta | \mathbf{y})$ was obtained by sampling from the product of the likelihood and the prior $p(\mathbf{y} | \theta) p(\theta)$ using the Metropolis-Hastings algorithm \cite{Gelman13, Kalos86}, a Markov-Chain Monte-Carlo (MCMC) sampling method \cite{Craiu14, Hastings70}. Thus, a distribution over the parameters of the device likely to produce the experimental J-V measurement may be determined, without an explicit model of this inverse dependence.

It was also found possible to decrease the variance of final posterior distributions over mobile ion vacancy parameters by including J-V characteristics measured from multiple voltage scan-rates. The set of characteristics $\mathbf{y}$ used in the method above was therefore the concatenation of two sets of characteristics for scan-rates of $0.1$V/s and $1.0$V/s. Previous studies \cite{Cave20} have shown that hysteresis is maximised at scan-rates of $0.1$V/s and $1.0$V/s and were therefore selected for this method to maximise the information on mobile vacancy dynamics within the likelihood. An analysis of this observation is provided in section 3 and describes how the method may be augmented to probe all regions of the mobile ion vacancy parameter space, enabling characterisation of an arbitrary device.

\begin{table}[h!]
\caption{\label{Table1}Device parameters $\theta$ and associated range of uniform prior distribution $p(\theta)$ for a TiO$_2$-MAPbI$_3$-Spiro PSC architecture. Parameter range represents plausible parameter values prior to any J-V measurement and may be tailored to a specific device if prior knowledge is available (see Section 3).}
\begin{indented}
\item[]\begin{tabular}{@{}l*{15}{@{\extracolsep{12pt plus12pt}}l}}
\br
Device parameter $\theta$ & Parameter range in $p(\theta)$ & Reference\\
\mr
\textbf{Perovskite layer} \\
Layer width $b_p$ & $150 - 600$ nm & \cite{Cave20} \\
Permittivity $\epsilon_p$ & $10 - 40 \epsilon_0$ Fm$^{-1}$ & \cite{Brivio14,Futscher19} \\
Absorption coefficient $\alpha$ & $4.0 \times 10^6 - 4.0 \times 10^7$ m$^{-1}$ & \cite{Loper15,Riquelme20} \\
Electron diffusion coefficient $D_n$ & $1.7 \times 10^{-6} - 1.7 \times 10^{-4}$ m$^2$s$^{-1}$ & \cite{Stranks13,Hill23} \\
Hole diffusion coefficient $D_p$ & $1.7 \times 10^{-6} - 1.7 \times 10^{-4}$ m$^2$s$^{-1}$ & \cite{Stranks13,Hill23} \\
Conduction band density of states $g_c$ & $10^{24} - 10^{25}$ m$^{-3}$ & \cite{Brivio14} \\
Valence band density of states $g_v$ & $10^{24} - 10^{25}$ m$^{-3}$ & \cite{Brivio14} \\
Mobile ion vacancy density $N_0$ & $10^{22} - 10^{26}$ m$^{-3}$ & \cite{Eames15, Walsh18} \\
Mobile ion diffusion coefficient $D_I$ & $10^{-17} - 10^{-12}$ m$^2$s$^{-1}$ & \cite{Richardson16,Cave20} \\
\textbf{ETL} \\
Doping density $d_E$ & $7.0 \times 10^{23} - 2.0 \times 10^{24}$ m$^{-3}$ & \cite{Schoonman81,Sellers11} \\
Layer width $b_E$ & $40 - 250$ nm & \cite{Riquelme20, Mukametkali23} \\
Permittivity $\epsilon_E$ & $7 - 35 \epsilon_0$ Fm$^{-1}$ & \cite{Stamate03,Guerrero14} \\
Electron diffusion coefficient $D_E$ & $10^{-8} - 5.7 \times 10^{-5}$ m$^2$s$^{-1}$ & \cite{Tiwana11,Krasienapibal14} \\
\textbf{HTL} \\
Doping density $d_H$ & $8.0 \times 10^{21} - 2.0 \times 10^{24}$ m$^{-3}$ & \cite{Abate14,Hochgesang22} \\
Layer width $b_H$ & $40 - 250$ nm & \cite{Luo19,Riquelme20} \\
Permittivity $\epsilon_H$ & $2 - 5 \epsilon_0$ Fm$^{-1}$ & \cite{Poplavskyy03,Rana11} \\
Hole diffusion coefficient $D_H$ & $10^{-8} - 10^{-5}$ m$^2$s$^{-1}$ & \cite{Leijtens13,Li17} \\
\br
\end{tabular}
\end{indented}
\end{table}

\section{Results and Discussion}
\begin{figure}
    \begin{subfigure}{.5\linewidth}
        \centering
        \includegraphics[width=0.8\linewidth]{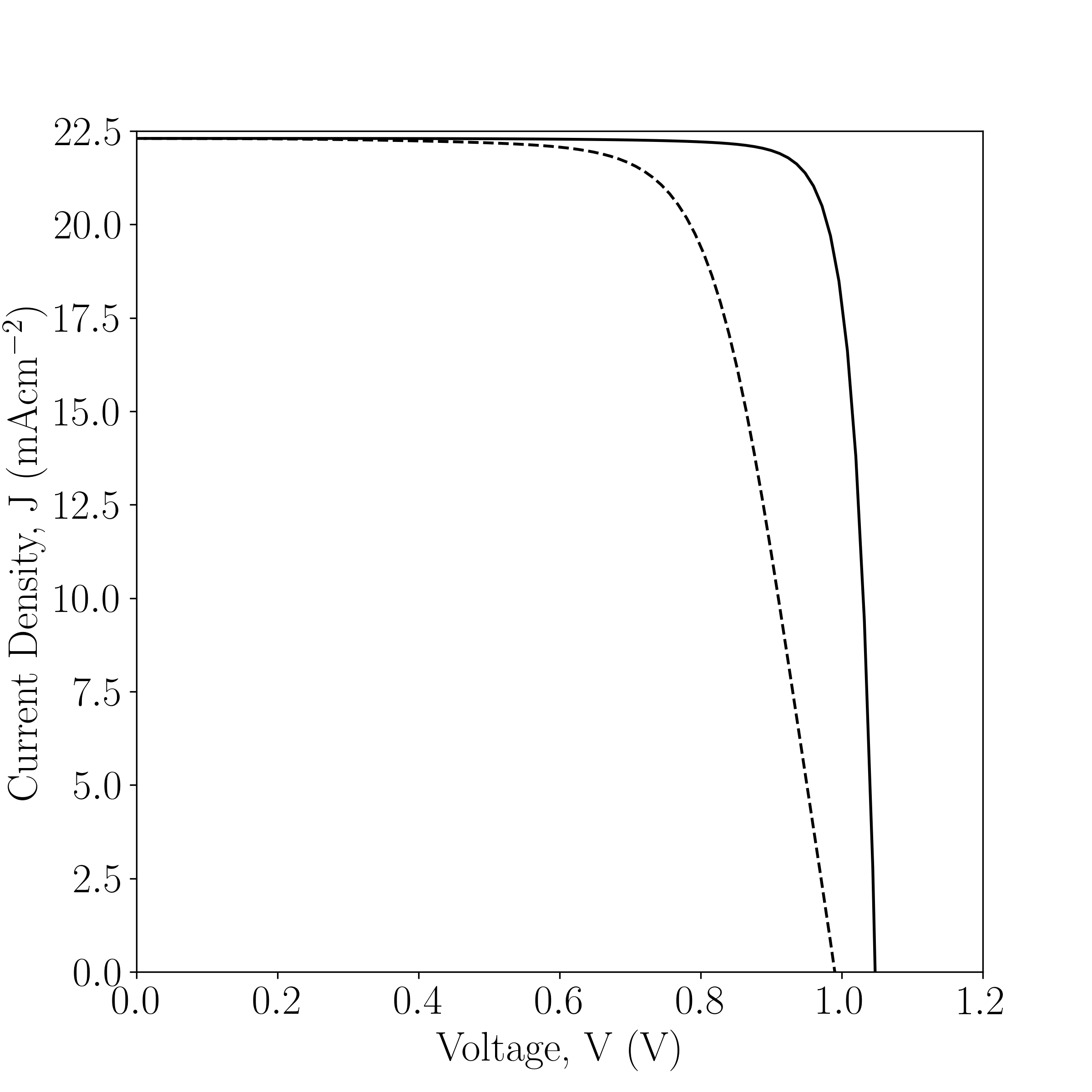}
        \caption{}
    \end{subfigure}%
    \begin{subfigure}{.5\linewidth}
        \centering
        \includegraphics[width=0.8\linewidth]{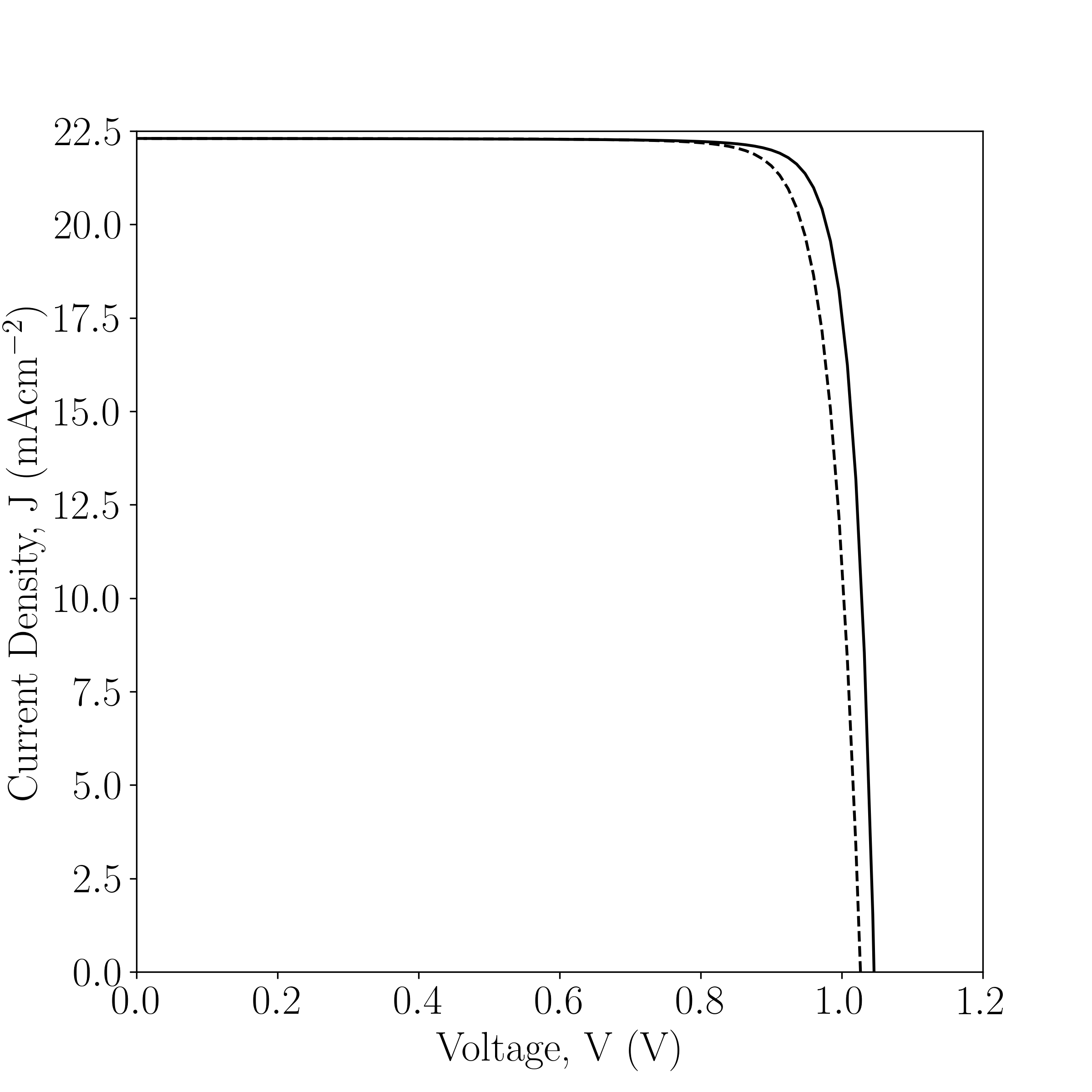}
        \caption{}
    \end{subfigure}\\[1ex]
    \begin{subfigure}{.5\linewidth}
        \centering
        \includegraphics[width=0.85\linewidth]{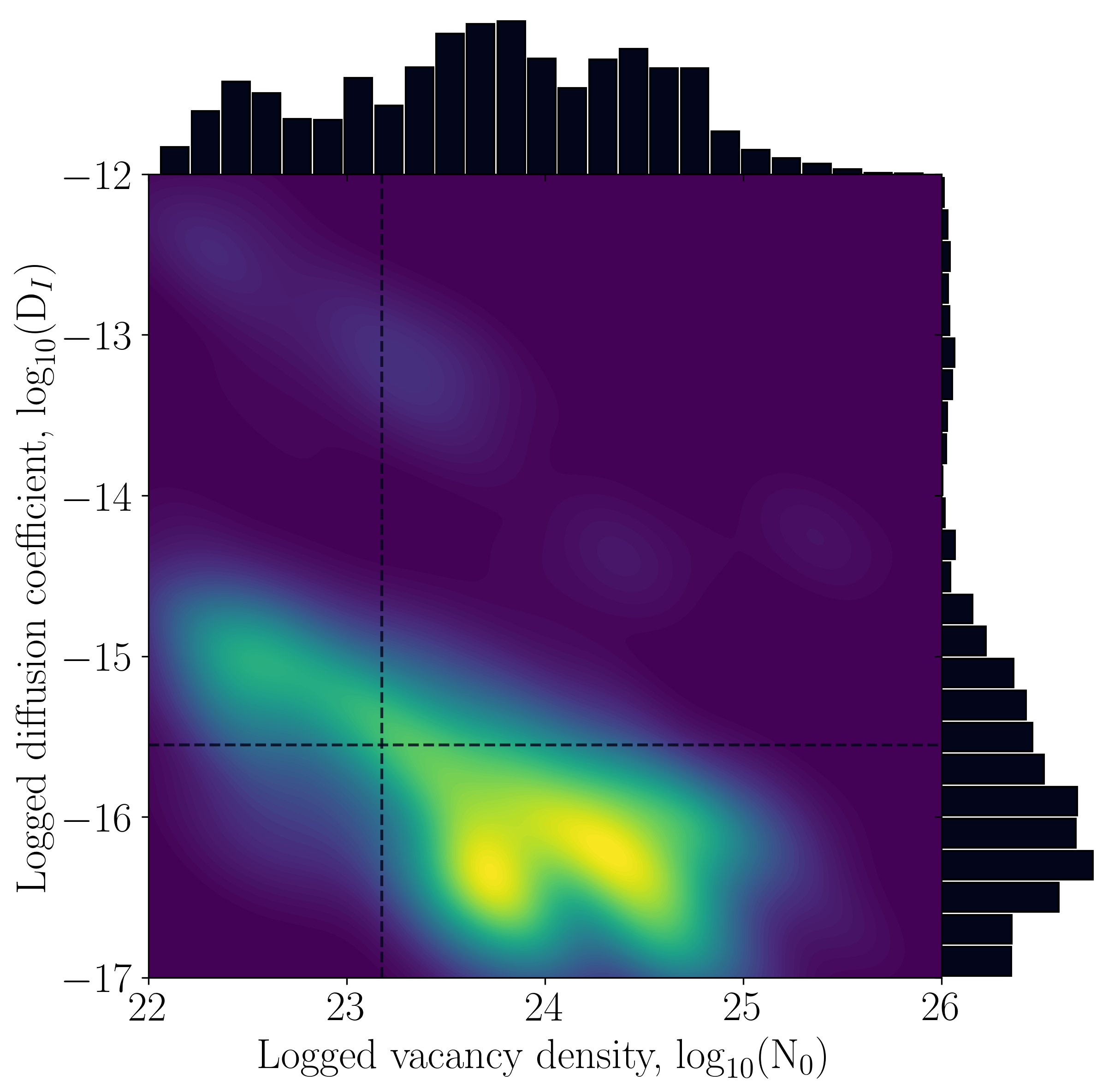}
        \caption{}
    \end{subfigure}%
    \begin{subfigure}{.5\linewidth}
        \centering
        \includegraphics[width=0.85\linewidth]{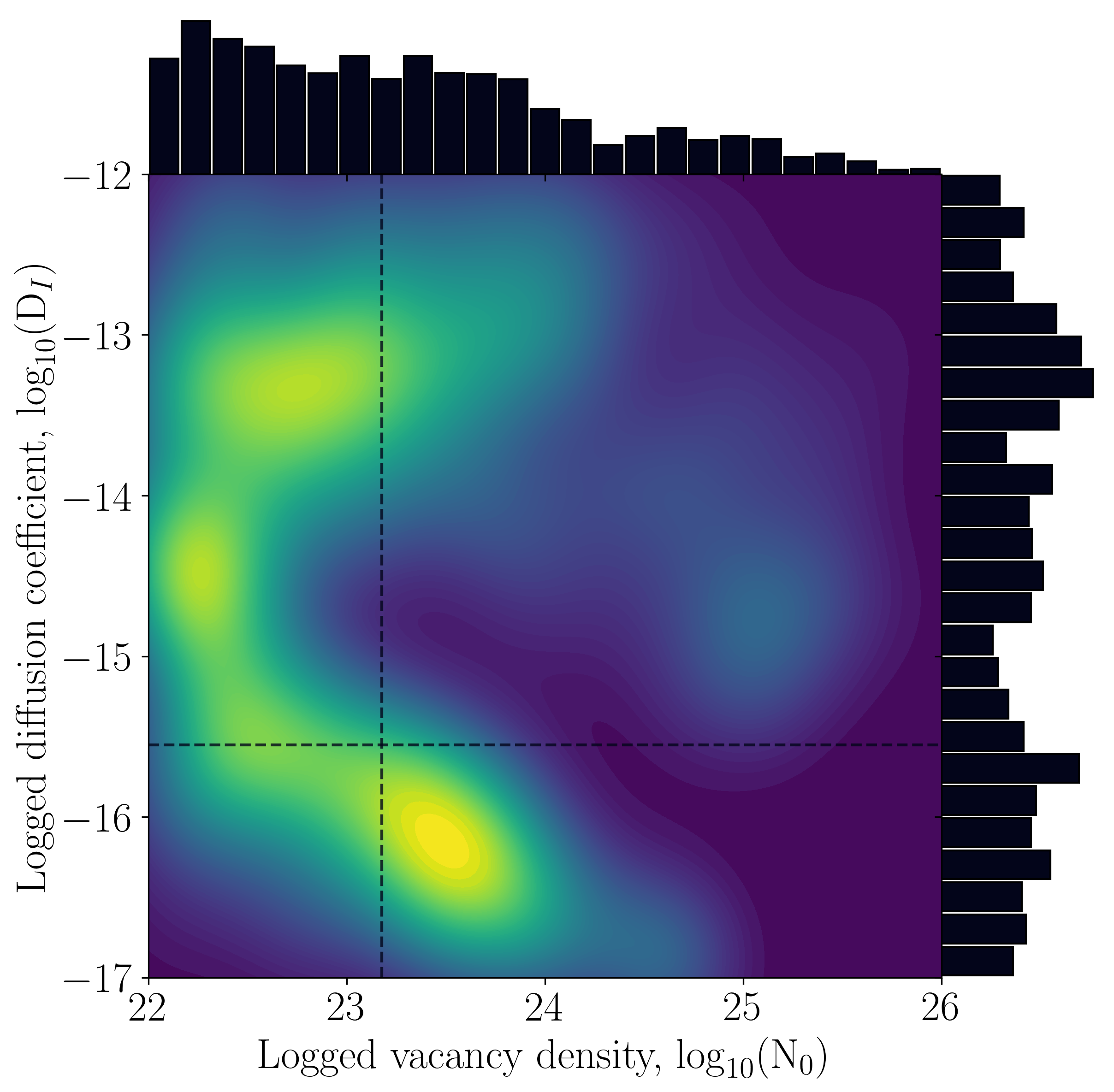}
        \caption{}
    \end{subfigure}\\[1ex]
    \begin{subfigure}{\linewidth}
        \centering
        \includegraphics[width=0.45\linewidth]{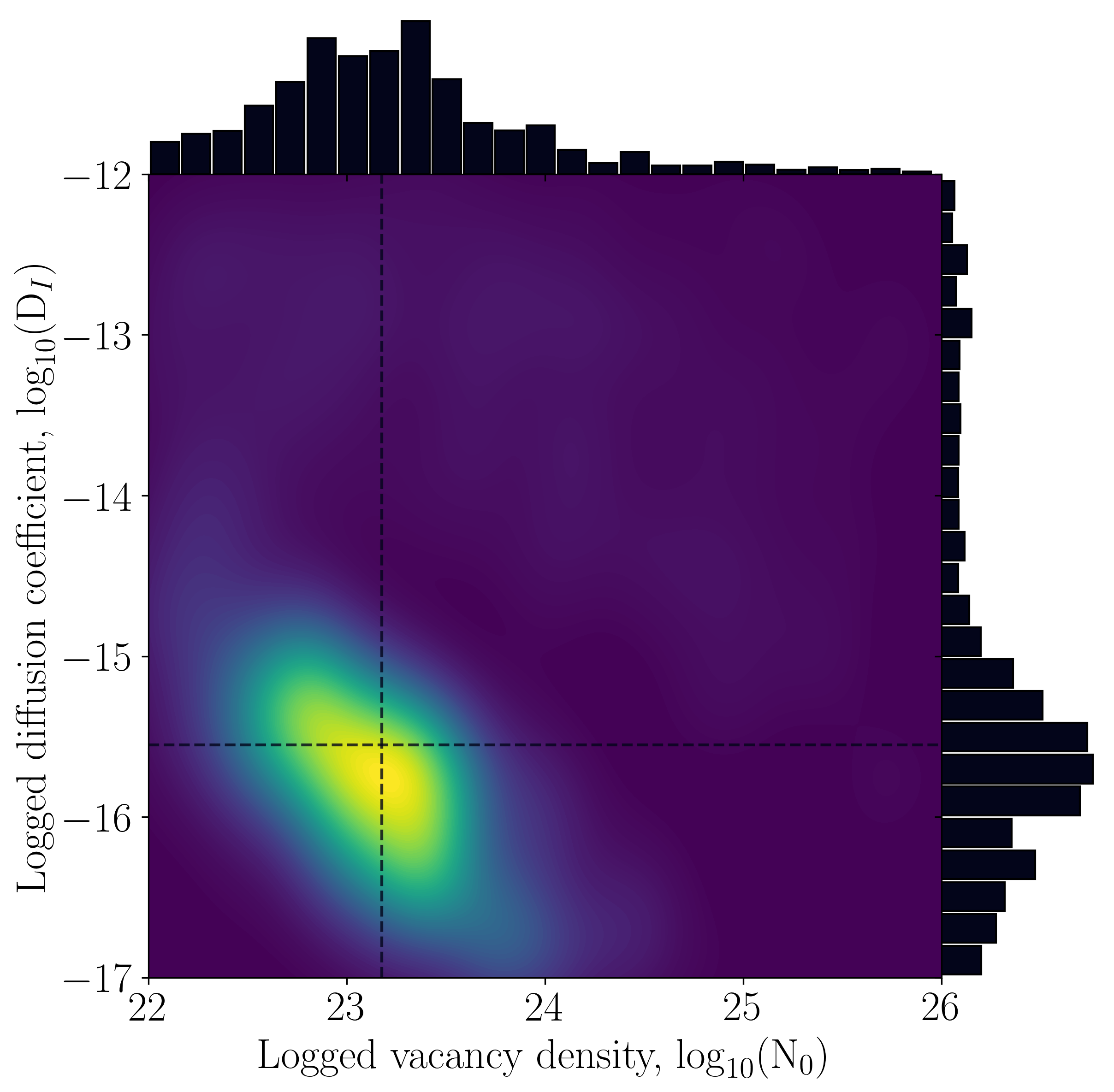}
        \caption{}
    \end{subfigure}
    \caption{Simulated J-V measurements (a) $0.1$mV/s, (b) $1.0$mV/s and associated posterior distributions shown in (c) and (d), respectively, where the likelihood $p(\mathbf{y} | \theta)$ has been specified by single scan-rate J-V characteristics. (e) Posterior distribution obtained with the likelihood specified by J-V characteristics obtained from both scan-rates. High posterior density indicated by yellow, fading to dark purple for zero density. True parameter values given by dashed reference lines.}
    \label{fig:main_posterior}
\end{figure}

To evaluate the method, a TiO$_2$-MAPbI$_3$-Spiro architecture was simulated by IonMonger and J-V measurements taken at $0.1$V/s and $1.0$V/s scan-rates. Using the BPE method presented, posterior distributions over mobile ion vacancy parameters, vacancy density $N_0$ and diffusion coefficient $D_I$, were determined (see Figure \ref{fig:main_posterior}). Specifically, $6$ Markov Chains were iterated and allowed to search the device parameter space for $500$ Metropolis-Hasting steps, resulting in $3000$ total device parameter samples. This process required $\sim 2.5$ hours of computation on a $32$ core Intel Xeon $2.40$GHz processor (2012), with IonMonger parallelised across all cores.

It can be seen from Figure \ref{fig:main_posterior} that the resulting posterior distribution obtained over $N_0$ and $D_I$ is a precise (i.e. non-zero probability density varying over $\sim 1$ order of magnitude) characterisation of the true mobile ion vacancy parameters associated with the observed J-V measurements, as simulated by IonMonger. The resulting distribution allows an understanding of internal mobile ion vacancy dynamics; specifically, whether poor/optimal device performance may be attributed to the number density of mobile ion vacancies (large $N_0$) and/or the vacancy diffusion coefficient $D_I$ is at a value likely to affect device performance at a given scan-rate. Additionally, posterior distributions were obtained from J-V characteristics taken at each scan-rate ($0.1$V/s and $1.0$V/s) individually at a temperature of 295K. The variance associated with the individual posteriors is significantly increased in comparison to the final posterior distribution obtained using J-V characteristics at both scan rates to specify the likelihood. This may be understood by considering that the posterior distribution obtained from both scan-rates must be determined from the overlap of the posterior distributions from individual scan-rates. The probability that a set of mobile ion parameters is associated with a set of J-V characteristics must now be determined from the probability the parameters are associated with J-V characteristics measured at $0.1$V/s \emph{and} $1.0$V/s. This overlap of the two likelihoods at each scan-rate therefore greatly reduces the probable mobile ion parameter space and reduces the uncertainty in characterisation.

Further, observed J-V characteristics exhibit a strong dependence on the scan-rate at which the device was measured \cite{Courtier19EES, Cave20}. Importantly, the matching of time-scales between ion diffusive motion and the rate at which the voltage is scanned determines whether J-V hysteresis is observed. Cave et al. \cite{Cave20} showed that by varying the scan-rate and locating the rate that causes the largest J-V hysteresis, a prediction of the mobile ion diffusion constant may be obtained. Similarly, here the scan-rate may be varied to 'probe' an arbitrary region of the ($N_0$, $D_I$) parameter space, with resulting J-V characteristics specifying the likelihood of the BPE method. As J-V hysteresis is strongly dependent on the dynamics of mobile ion vacancies, maximising observed J-V hysteresis was found to provide the highest characterisation precision within the BPE method (see Figure \ref{fig:main_posterior}(c) and (d)). This freedom to increase characterisation precision in any region of mobile ion parameter space by varying a single parameter (scan-rate), allows for characterisation of an arbitrary device with arbitrary values of $N_0$ and $D_I$.

\begin{figure}
    \centering
    \includegraphics[scale=0.8]{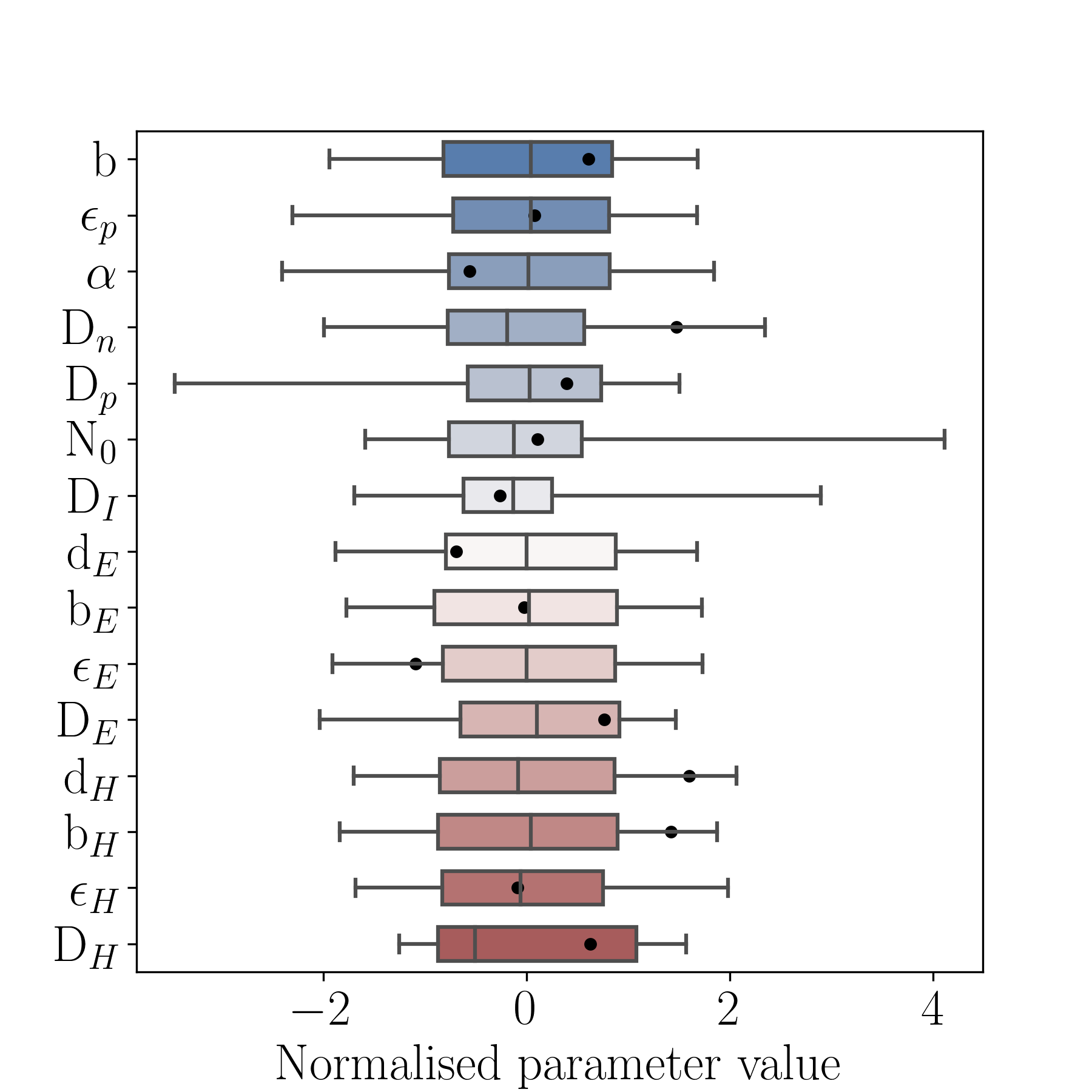}
    \caption{Boxplot over posterior parameter samples obtained via BPE method for each normalised physical device parameter (see Table 1). Lower, median and upper quartiles shown by vertical lines of central box. Whiskers are inclusive of extremal parameter samples and therefore indicate the extent of the Markov-Chain search over prior uniform range for each parameter. Black dots show true device parameters associated with the simulated J-V curves given in Figure \ref{fig:main_posterior}.}
    \label{fig:boxplot}
\end{figure}

As shown in Table 1, all physical device parameters were allowed to vary within a uniform prior range. Consequently, characterisation of all device parameters was obtained via BPE and a visualisation of the posterior distributions across all parameters is shown in Figure \ref{fig:boxplot}. Figure \ref{fig:boxplot} shows that parameters $N_0$ and $D_I$ are characterised with the least variance; the layer width $b$, permittivity of perovskite layer $\epsilon_p$ and absorption coefficient $\alpha$ are also consistently well characterised but with larger uncertainties. The plausible parameter ranges over other physical device parameters have been considerably reduced, with varying degrees of characterisation precision. The method presented has been optimised to characterise $N_0$ and $D_I$ by specifying the likelihood with measured J-V characteristics that are strongly dependent on these parameters (e.g. $\mathrm{P}_\mathrm{max}^\mathrm{f}$). However, it is known that device performance, as characterised by J-V measurements, is also dependent on other physical parameters such as the doping densities of the transport layers \cite{Courtier19EES}. It is therefore important that all device parameters were allowed to vary and affect performance (J-V characteristics) to ensure that any under-performance is not wrongly attributed to the effect of ion motion. In principle, the precision of characterisation over any device parameter may be improved by incorporating further device measurements within the likelihood that are known to be strongly dependent on that device parameter. For example, pulse J-V measurements have been shown to allow for characterisation of PSCs while minimising the effect of ion migration \cite{Hill23}. By specifying the likelihood of the BPE method with characteristics measured by the pulse J-V technique, the influence of mobile ion vacancy parameters on the measurement can be minimised. Other device parameters may therefore be more precisely characterised as their effect on performance is amplified. Here, the J-V characteristics selected to specify the likelihood lead to high characterisation precision for mobile ion parameters $N_0$ and $D_I$ and lower characterisation precision for transport layer parameters. This indicates that the selected J-V characteristics are less dependent on changes in the transport layer parameters than mobile ion parameters. Indeed, high characterisation precision of $N_0$ and $D_I$ has been obtained despite poorer characterisation of the transport layer parameters. This suggests the potential to de-couple the influence of mobile ion parameters on performance by selection of a relevant set of performance characteristics (here, J-V characteristics) strongly dependent on the mobile ion parameters and weakly dependent on other device parameters. Again this reasoning may be extended to the characterisation any device parameter via a BPE method.

\begin{figure}
    \begin{subfigure}{.5\linewidth}
        \centering
        \includegraphics[width=0.80\linewidth]{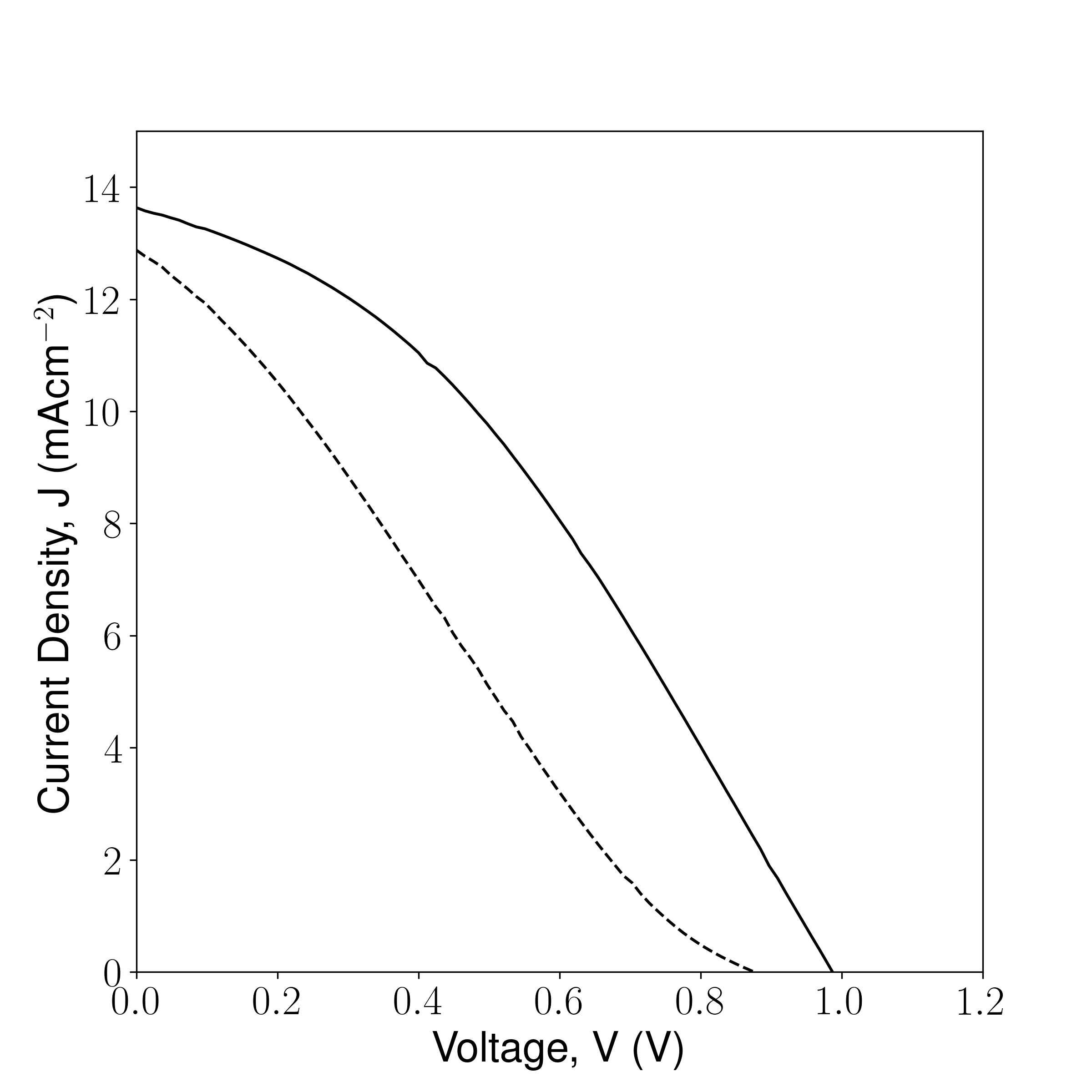}
        \caption{}
    \end{subfigure}%
    \begin{subfigure}{.5\linewidth}
        \centering
        \includegraphics[width=0.80\linewidth]{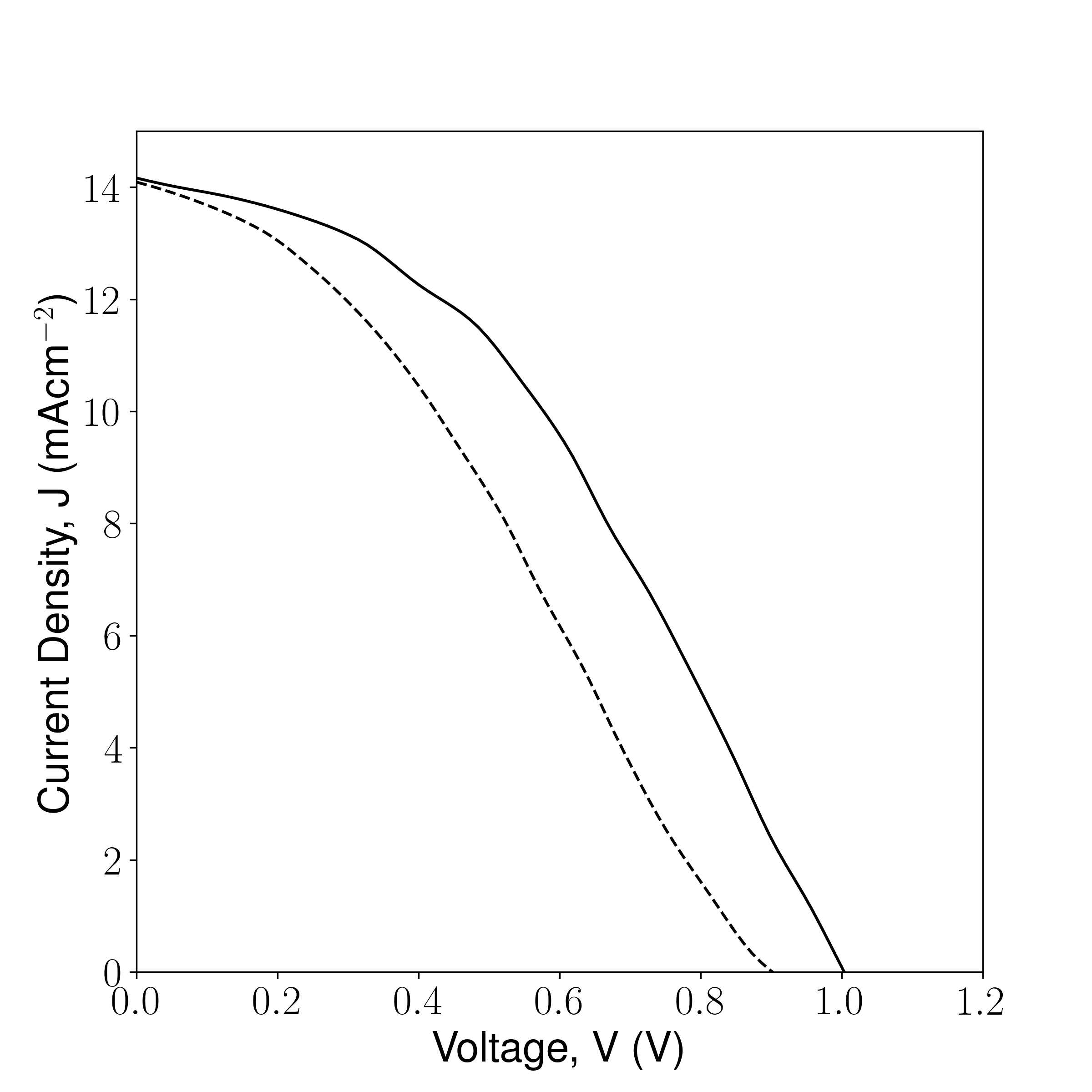}
        \caption{}
    \end{subfigure}\\[1ex]
    \begin{subfigure}{\linewidth}
        \centering
        \includegraphics[width=0.45\linewidth]{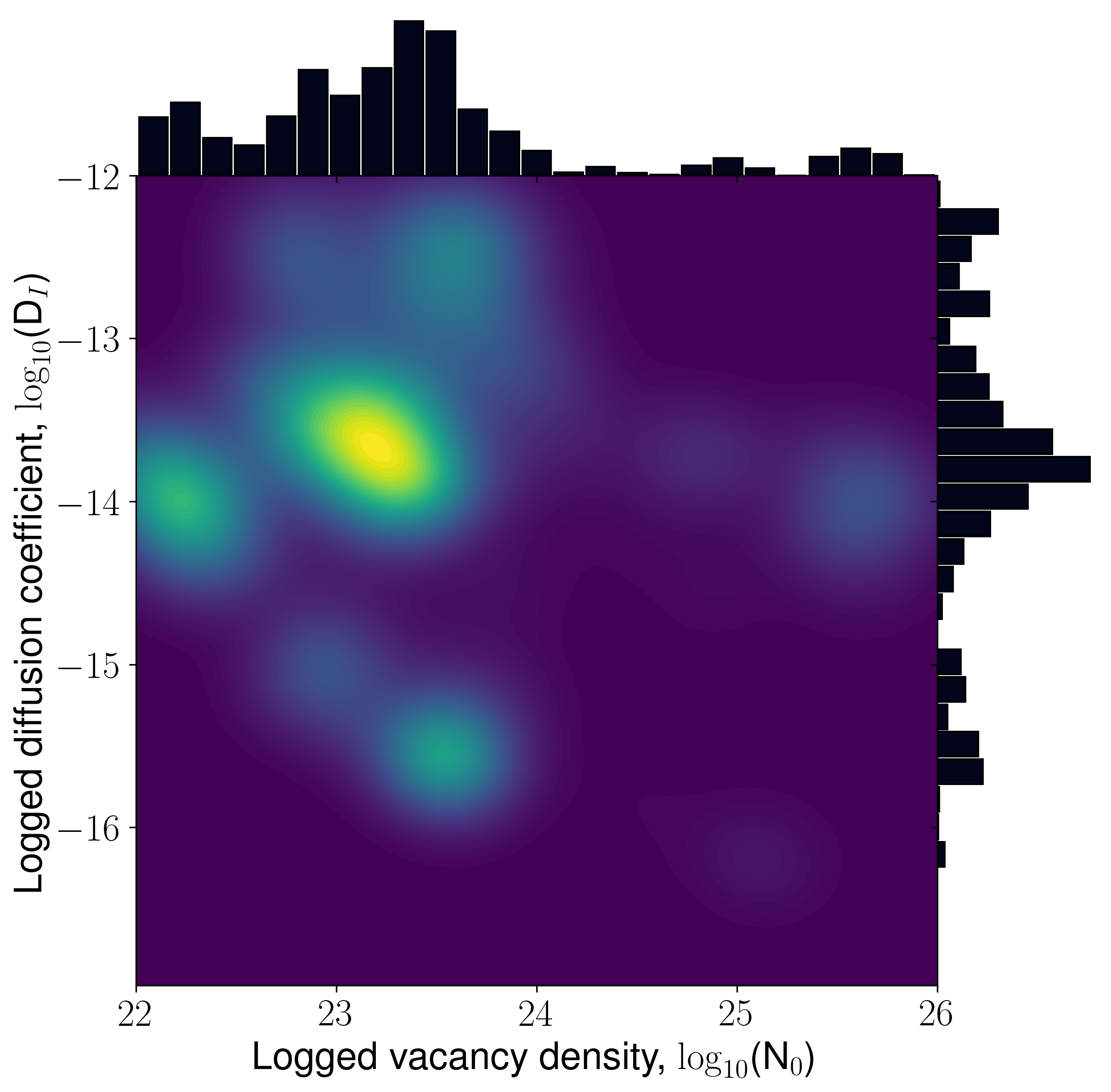}
        \caption{}
    \end{subfigure}
    \caption{Experimental J-V measurements at scan rates of (a) $0.178$V/s and (b) $1.0$V/s. (c) Associated posterior distribution over mobile ion vacancy parameters, $N_0$ and $D_I$. High posterior density indicated by yellow, fading to dark purple for zero density.}
    \label{fig:exp_posterior}
\end{figure}

The method was tested experimentally with J-V measurements on a TiO$_2$-MAPbI$_3$-Spiro architecture from Cave et al. \cite{Cave20} where the device was measured at scan-rates of $0.178$V/s and $1.0$V/s at 295K. A posterior distribution over mobile ion parameters $N_0$ and $D_I$ was obtained using the BPE method and the result can be seen in Figure \ref{fig:exp_posterior} as well as the two J-V measurements taken by Cave et al. \cite{Cave20} Figure \ref{fig:exp_posterior} shows how $N_0$ and $D_I$ are deduced to have median values of $4.9 \times 10^{23}$m$^{-3}$ and $3.5 \times 10^{-14}$m$^2$s$^{-1}$, respectively. Assuming the ion diffusion coefficient $D_I$ has an Arrhenius dependence on temperature $T$, $D_I=D_{I,\infty}\exp[-E_a/(k_BT)]$, where the high temperature diffusion coefficient $D_{I,\infty} = 3\times 10^{-8}$m$^2$s$^{-1}$ for a vacancy hop attempt frequency of $1$THz \cite{Cave20}. The vacancy hop activation energy $E_a$ may be predicted from this relationship and was calculated to be $0.35$ eV. This prediction lies within experimentally obtained activation energies of 0.36–0.43 eV. \cite{Tammireddy22} Cave et al \cite{Cave20} predicted 0.41 eV for $E_a$ in a PSC with HTL, ETL of spiro, TiO$_2$ respectively and the DFT prediction of $0.37$eV ($0.51$eV) for tetragonal (cubic) MAPbI$_3$,  allowing for a reaction enthalpy of 0.07 eV \cite{CavePhDthesis18}, and noting that the cubic-to-tetragonal phase transition in MAPI occurs around 54 °C,  327K \cite{Patru21}. However, individual device characterisation, and this method, assume that mobile ion vacancy parameters vary between devices as a result of fabrication and device history and DFT predictions for $E_a$ are approximate. The single DFT prediction for all TiO$_2$-MAPbI$_3$-Spiro architectures assumes a level of uniformity across devices that oversimplifies the resulting mobile ion dynamics in PSCs. Using the BPE method, the prediction for the activation energy of mobile ion vacancy diffusion is device specific. This therefore allows for an analysis to be constructed, correlating fabrication methods with resulting mobile ion vacancy dynamics.

Additionally, an analysis of the assumptions of the numerical simulation model, IonMonger, leads to further potential insight. Specifically, IonMonger assumes that all ion vacancies are mobile - an assumption which is known not to hold in fabricated PSCs due to, for example, grain boundary dominated migration \cite{Shao14}. The resulting posterior over vacancy density $N_0$ predicted by this BPE method is therefore a quantification of the density of \emph{mobile} vacancies. The number of total ion vacancies may be larger due to immobile ion vacancies. For example, DFT calculations predict an iodide vacancy density for cubic MAPbI$_3$ to be $1.6 \times 10^{25}$m$^{-3}$ at room temperature \cite{Eames15, Walsh18}; whereas Figure \ref{fig:exp_posterior} predicts a iodide vacancy density of $4.9 \times 10^{23}$m$^{-3}$. This discrepancy may be due to either a lower number of total ion vacancies resulting from fabrication or a lower number of \emph{mobile} ion vacancies, with migration heavily limited by the density of grain boundaries.

\begin{figure}[b]
\centering
\begin{subfigure}{.5\textwidth}
  \centering
  \includegraphics[width=\linewidth]{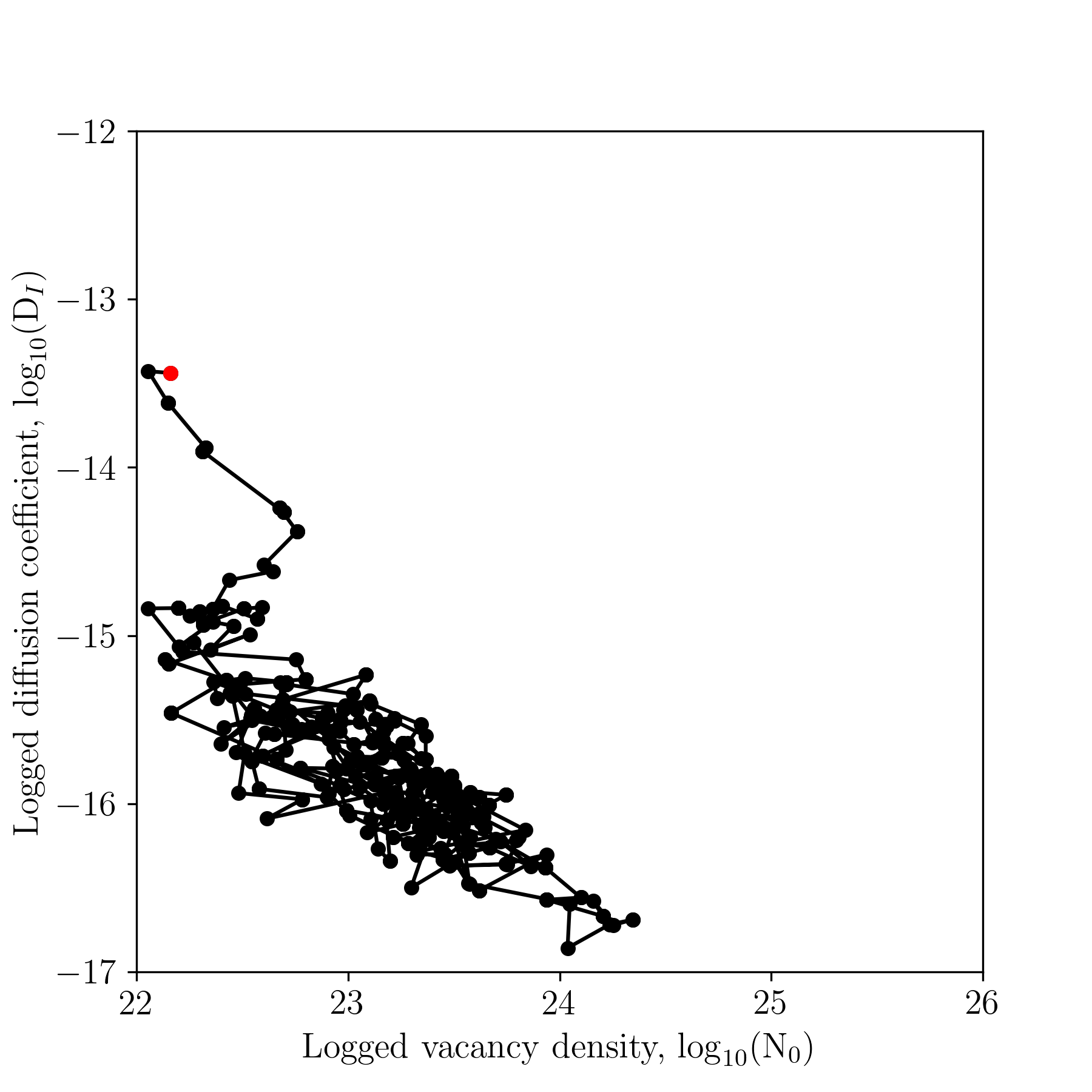}
\end{subfigure}%
\begin{subfigure}{.5\textwidth}
  \centering
  \includegraphics[width=\linewidth]{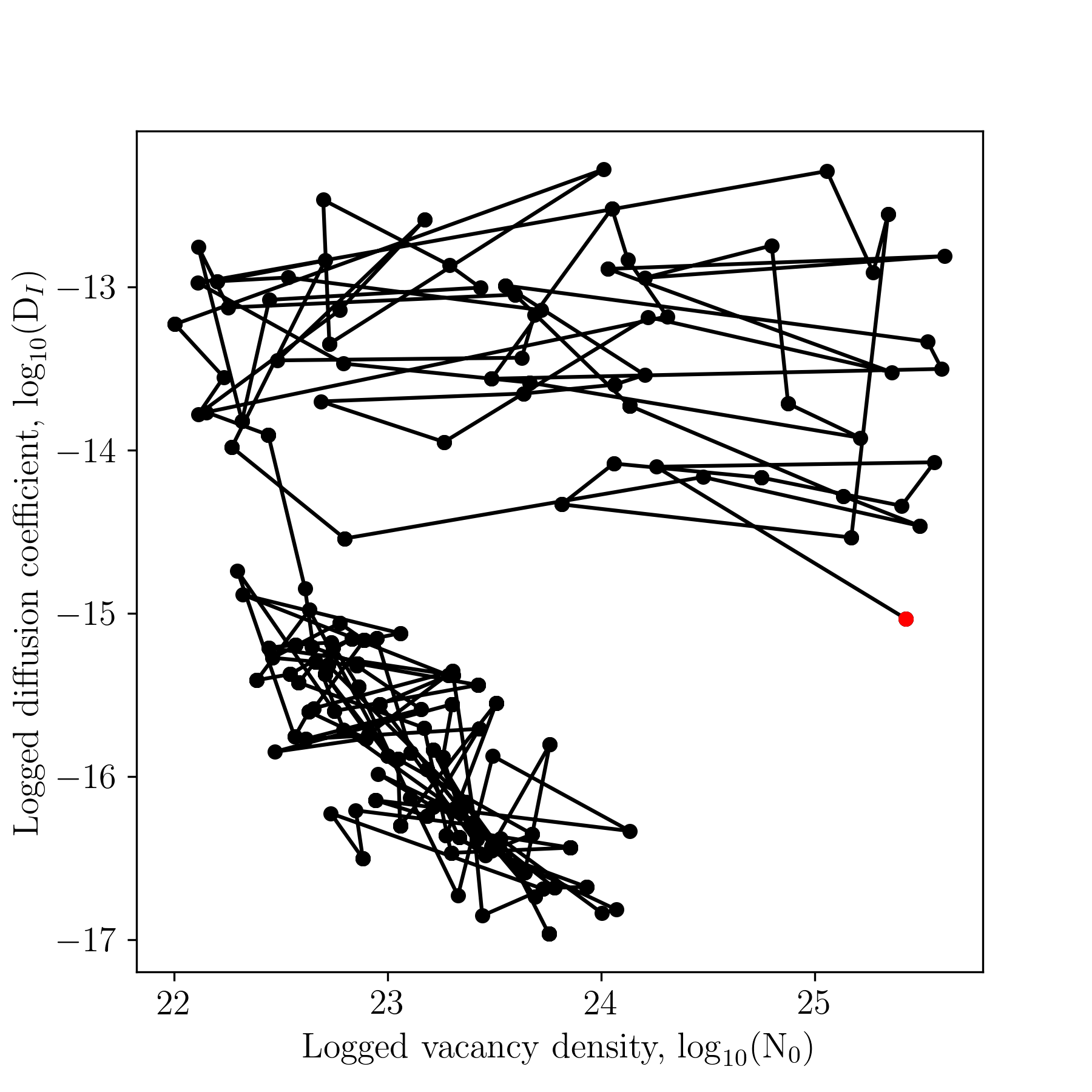}
\end{subfigure}
\caption{Example Markov Chains for jumping distribution standard deviations given by $q=1\%$ (left) and $q=4\%$ (right). Red points indicate initial randomly sampled starting point of the Markov Chain. The resulting posterior ($q=1\%$) is given in Figure \ref{fig:main_posterior}(e).}
\label{fig:markov_chains}
\end{figure}

The efficiency of the BPE method, and most importantly the efficiency of the Metropolis-Hasting algorithm, was observed to be strongly dependent on jumping distribution variance (see Figure \ref{fig:markov_chains}). Within the Metropolis-Hastings algorithm, each Markov chain proposes a new position in the parameter space by sampling a point in this space from a jumping distribution centered on the current point. Here, the jumping distribution was specified by a multivariate Gaussian distribution with the standard distribution in each parameter given by a fixed percentage, $q$, of the prior distribution range (see Table 1). For large values of $q$, the algorithm takes large steps around the parameter space, potentially finding a region of the space with high-likelihood more quickly. Yet, due to the large jumping distribution variance, the acceptance rate of proposed states will be low and the algorithm is unlikely to remain precisely in this region. Conversely, if the value of $q$ is too low the acceptance rate will increase but the algorithm may take too long to find a region of the space with high-likelihood. Optimal scaling theory provides a prediction of the asymptotically optimal acceptance rate of the Metropolis-Hasting algorithm for a multi-dimensional parameter space to be $23.4\%$ \cite{Roberts01}. The variance of the jumping distribution was therefore tailored to achieve this acceptance rate on average, which was found to be for $q\sim 1\%$.

Future work may look to improve the efficiency of this method via superior MCMC algorithms such as Hamiltonian Monte Carlo (HMC) \cite{Betancourt17} or No-U-Turn Samplers (NUTS) \cite{Hoffman14}. These methods utilise the derivative of the likelihood to efficiently search the parameter space, requiring fewer iterations/evaluations of the posterior and therefore, in this application, fewer device model queries. Active learning methods \cite{Tong01}, may also be investigated. Active learning methods look to specify the set of device parameters with which to query the drift-diffusion PSC model such that the decrease in entropy of the posterior distribution is maximized, potentially resulting in high query/data efficiency.

Application specific efficiency improvements may also be obtained by constraining the device parameter space with prior device knowledge. In the examples presented here, prior distributions over device parameters were uninformative uniform distributions over the full range of plausible values for MAPbI$_3$. Reducing the parameter space over which the method is performed would decrease the number of MCMC iterations required and may be achieved by application-specific informative prior distributions. For example, the prior distribution over perovskite layer width ranged from 400nm-900nm; however, the layer width is commonly characterized previously to J-V measurement. The prior distribution over this parameter may therefore be specified by a normal distribution centered on the measured value with variance given by the measurement uncertainty. This may be extended to any number of device parameters that have previously been characterized, greatly decreasing the searched parameter space.

\section*{Conclusions}
 A method for fast ($\sim$ hours) reverse modelling, mapping from J-V device measurements to device and materials parameters, has been presented. We have demonstrated that mobile halide ion parameters can be deduced from room temperature J-V measurements \emph{only}, requiring no further experimental protocols. In our method, Bayesian Parameter Estimation is coupled with a drift-diffusion PSC model, to determine posterior distributions over mobile ion parameters which were found to precisely characterise the ion vacancy density $N_0$ and diffusion coefficient $D_I$ for both simulated and fabricated devices. With this method, an understanding of the internal dynamics of mobile ions, a phenomenon known to cause J-V hysteresis and significantly hinder optimal charge-transport, may be obtained for a \emph{specific} device. This understanding allows fabrication techniques and other potential device optimisations to be correlated with the resulting internal dynamics of mobile ion vacancies. Further, by performing characterisation over all physical device parameters, the influence of mobile ion vacancies on performance may be decoupled from the effect of other physical parameters. The relative influence of physical processes within the device may therefore be established.

The method presented here can also be applied in a digital twin-style setting, where a virtual model of a laboratory device is created and continuously updated to reflect the current output of the device. The digital twin would use our drift diffusion code combined with Bayesian Parameter Estimation to find values for the drift-diffusion model input parameters that reproduce measurements of the PSC being studied, such as its current-voltage (J-V) characteristics. If the device changes over time, e.g. due to
degradation, this method would be able to identify the causes of changes in device performance and track these over time. This approach could be of value, not only in current efforts to improve PSC by helping to determine underlying physical processes, but also in the future by employing it in operational settings, where methods for monitoring and optimising devices would be needed.

High-performance computing (HPC) may be leveraged to further reduce the characterisation time, potentially allowing for full device characterisation in minutes, requiring only J-V measurements as input. Further methodological improvements may also be investigated such as improved MCMC sampling and active learning methods.

\section*{Acknowledgments}
We would like to thank James Cave for permission to use the data of the J-V measurements from his PhD thesis \cite{CavePhDthesis18} reported in Figure 4 of this paper.  SM and JL thank the UK Engineering and Physical Sciences Research Council (EPSRC) for respectively a summer bursary from Computational Collaboration Project No 5 (CCP5) and a doctoral training partnership studentship. MVC was supported by the University of Bath Institute for Sustainability, EP/L016354/1.

\printbibliography
\end{document}